\documentstyle[aps]{revtex}
\begin{document}
\title{Finiteness following from underlying theory: a natural strategy}
\draft
\author{Jifeng Yang}
\address{School of Management,Fudan University, Shanghai, 200433, P. R.
China}
\maketitle

\begin{abstract}
A tentative proposal is demonstrated that there is a natural strategy to get
rid of unphysical (UV) infinities in QFTs if one adopts the modern standard
point of view that a fundamental theory that is complete and well-defined in
all respects underlies the QFTs. This simple strategy works in principle for
any interaction model and space-time dimension. It provides a physical
rationality behind the UV divergence and the conventional renormalization
programs and improves the latter in several important aspects.
\end{abstract}


% repeat the \author\address pair as needed

%\date{\today}

\section{Introduction}

Now, it has become a standard point of view that a fundamental theory (well
defined for the extremely high energy end) underlies the present QFTs that
are in fact low energy (LE) effective theories for the phenomena in LE
ranges \cite{wein}. But as far as the author knows, we are still lacking a
formulation that can yield finite results in a natural way that fully makes
use of the standard point of view. A new strategy is proposed in Ref. \cite
{YYY} that indicates the power of the standard point of view if one uses it
appropriately.

To focus on the UV problem, we will assume from now on that there were no
unphysical IR singularity in the LE models in our discussions or we have
already had an IR regular formulation for the LE QFTs. (We will discuss in
our future works about the IR structure's contribution to the whole
formulation--it should be included to arrive at a totally satisfying
formulation, especially for QCD-like theories where the IR singularity is
rather severe and affects the theories' predictions \cite{Bigi}).

It is convenient to employ a generating functional formalism \cite{Shirkov}
or a path integral formalism to assemble the Green functions for the LE
sectors of the underlying theory. It is natural to expect that the
well-defined path integral for an effective sector take the following form,
%%EQ(1)
\FL
\begin{equation}
Z_{\{\sigma\}}(\{J^{i}\})= \int D{\mu}(\phi^{i}_{\{\sigma\}}) \exp
\{iS(\phi^{i}_{\{\sigma\}};\{J_{i}\}; \{\sigma\})\}
\end{equation}
where $\{\sigma\}$ are the fundamental parameters (some fundamental
constants, probably including the gravitation constant) from the underlying
theory and $\{J^{i}\}$ are the external sources for LE sectors. The
'elementary fields' $(\phi^{i}_{\{\sigma\}})$ in the LE sectors are here
appended by the underlying parameters to indicate that they are in fact
effective ones. It is easy to see that for different LE physics, the LE
limit operation may act upon sets of underlying parameters that differ in
part.

Within the path integral formalism, we can easily see that in the LE limit
(denoted by ${\bf L}_{\{\sigma \}}$) %%%Eq(2)
\FL
\begin{eqnarray}
&&Z^0(\{J^i\};\overline{\cdots })\equiv {\bf L}_{\{\sigma \}}\int D{\mu }%
(\phi _{\{\sigma \}}^i)\exp \{iS(\phi _{\{\sigma \}}^i;\{J_i\};\{\sigma
\})\}
\\
&\neq &\int D{\mu }(\phi ^i)\exp \{iS(\phi ^i;\{J_i\})\},
\end{eqnarray}
where the symbols not appended by the underlying parameters refer to the
constants and field parameters given by the present QFTs (i.e., the LE limit
has been applied on these 'effective' fields or objects). Thus, generally
speaking, we can not let the operation of the LE limit cross the summation
over the paths or the intermediate states. In other words, the LE limit
operation and the summation over intermediate states {\it do not commute},
%%Eq.(4)
\FL
\begin{equation}
\Delta \equiv [{\bf L}_{\{\sigma \}},\sum_{\{paths\}}]\neq 0.
\end{equation}
If one ignores such a subtle fact, the resulting formulations would be ill
defined and needs regularizations that introduce unjustified artificial
substitutes of the true underlying structures, which will in turn lead to
divergences may appear and subtraction is needed. This is our opinion about
the appearance of the unphysical infinities in some conventional QFT
formulations.

In the path integral formalism it is immediate to see that the spectra given
by the conventional Hamiltonian models would differ from true ones given by
the underlying theory, especially in the UV regions. The deviation is
signaled by the ill-definedness or UV divergence in the conventional QFTs.
More severe UV divergence implies more severe deviation.

What we are trying to present in the following is that if one starts merely
with the existence of the underlying theory and $\{\sigma\}$ (without
knowing the details), there is a simple strategy to calculate the amplitudes
wanted without introducing any {\it ad hoc } regularization or cutoff that
leads to UV divergence. But, due to the lacking of the true underlying
structures, there {\it must remain} in our approach certain ambiguities
signaling the missing of the underlying structures, which are to be fixed in
principle by phenomenology and experiments.

This paper is organized in the following way: We will first exemplify our
strategy in Feynman graph language for the one-loop case in section II. The
treatment of the multi-loop cases is given in section III. Then we discuss
some nonperturbative examples in section IV to further illustrate our
proposal. Finally we summarize our presentation in Section V. We should note
in advance that here we do not claim a satisfactory, systematic and final
presentation of a new approach, instead we just put forward an alternative
strategy for dealing with ill-definedness.

\section{How Can UV Finite Results be Derived}

From our discussion in section I, we see that the Hamiltonians (and hence
the propagators and vertices) in their present forms are the LE limits of
the ones characterized by $\{\sigma \}$. To interest most readers we will
exemplify our strategy in the perturbative Feynman graph language (we remind
that our strategy is definitely not confined to the perturbative case and we
will exhibit the use of the strategy in some nonperturbative cases in
section IV). Thus, according to our postulate, the conventional ill-defined
(or divergent) Feynman amplitudes (FAs) are results of illegitimate
operation order with the LE limit operation and the internal momenta
integration. In formula, if the integrand $f(\{Q_i\},\{p_j\},\{m_k\})$ (of
an ill-defined FA) corresponds to the integrand $\bar{f}(\{Q_i\},\{p_j\},%
\{m_k\};\{{\sigma }_l\})$ given by the underlying theory with $%
\{Q_i\},\{p_j\},\{m_k\}$ and $\{{\sigma }_l\}$ being respectively loop
momenta, external momenta, masses and the fundamental parameters in the
underlying theory, then %Eq.(5)
\FL
\begin{eqnarray}
&&\Gamma ^0(\{p_j\},\{m_k\};\{\bar{c}\})={\bf L}_{\{\sigma \}}\overline{%
\Gamma }(\{p_j\},\{m_k\};\{\sigma _l\})={\bf L}_{\{\sigma \}}\int
\prod_id^nQ_i\bar{f}(\{Q_i\},\{p_j\},\{m_k\};\{\sigma _l\}) \\
&\neq &\int \prod_id^nQ_i{\bf L}_{\{\sigma \}}\bar{f}(\{Q_i\},\{p_j\},\{m_k%
\};\{\sigma _l\})=\int \prod_id^nQ_if(\{Q_i\},\{p_j\},\{m_k\}),
\end{eqnarray}
where $\{\bar{c}\}$ denotes the definite constants arising from the LE limit
operation. $\Gamma ^0$ and $\overline{\Gamma }$ are well-defined (finite)
but $\int \prod_id^nQ_if(\{Q_i\},\{p_j\},\{m_k\})$ is ill defined. That
means, the commutator %%Eq.(7)
\FL
\begin{equation}
\delta _{\{\sigma \}}=\left[ {\bf L}_{\{\sigma \}},\int \prod_id^nQ_i\right]
\end{equation}
only vanishes identically for convergent (i.e., well-defined) FAs, otherwise
we encounter divergence or ill-definedness in FAs. The deviation of the
effective formalism is not detected by the convergent FAs, or these
amplitudes are well defined in the present QFTs. This is an extremely
important fact for our purpose in the following.

As the underlying theory or the amplitudes $\bar{f}(...;\{\sigma_{l}\})$ are
unavailable by now, we have to find a way to approach the truth--the $%
\Gamma^{0}(\{p_{j}\},\{m_{k}\})$'s--from our present partial knowledge about
the LE sectors of the underlying quantum theory --the present form of the
Hamiltonians or Lagrangians as LE sectors of the underlying theory.

Here is our strategy for extracting finite results out of effective
formulations of QFT: (1) First we try to perform certain {\it legitimate}
operations (say, $\Omega $) on the objects {\it so that} the LE limit
operation commutes with the summation on the resulted objects if they do not
commute on the original objects. (2) Then we can safely perform the
intermediate states summation (e.g., loop integration) on the resulted
objects. (3) At last, we perform the inverse operations (say, $\Omega
^{-1}$%
, which should also be legitimate) to go back and the final expressions
should be (UV) finite by construction but probably ambiguous at the meantime
due to our lack of knowledge about the underlying structures. The
ambiguities should be fixed from the phenomenological and experimental
inputs as mentioned above. The operations ($\Omega $'s) and their inverse
are of the main concern in our strategy and they would often be operations
with respect to the parameters external to the intermediate states since the
objects of interests are expressed in terms of these parameters after all.

The more general framework effecting the natural strategy proposed here will
be the subject of our future investigations. Here, we only show how the
strategy works in a simple way for the Feynman graph approach, i.e., we
exemplify our simple strategy in the Feynman graph language. First we show
that the following important relation holds for 1-loop case ill-defined FAs
(c.f. Eq.(5) for 1-loop case) %%eq.(8)
\FL
\begin{equation}
\int d^{n}Q \left ({\partial}_{p_{j}} \right )^{\omega}
f(Q,\{p_{j}\},\{m_{k}\})= \left ( {\partial}_{p_{j}} \right )^{\omega}
\Gamma^{0} (\{p_{j}\},\{m_{k}\}),
\end{equation}
with $\omega-1$ being the usual superficial divergence degree of $\int
d^{n}Q f (Q,\{p_{j}\},\{m_{k}\})$ so that the left hand side of Eq.(8)
exists (finite), $\left ({\partial}_{p_{j}} \right
)^{\omega} $ denoting
the differentiation's with respect to the external parameters $\{p_{j}\}$'s
of the amplitude and $\Gamma^{0}(...)$ is the LE limit of the amplitude
calculated in the underlying theory (i.e., the internal momentum integration
is performed first). It is easy to see that the operation $\left
({\partial}%
_{p_{j}} \right )^{\omega}$ leads to convergent graphs or objects.

The proof is very simple, since %Eq(9)
\FL
\begin{eqnarray}
&&\int d^{n}Q \left ({\partial}_{p_{j}} \right )^{\omega} f
(Q,\{p_{j}\},\{m_{k}\})= \int d^{n}Q \left ({\partial}_{p_{j}}
\right)^{\omega} {\bf L}_{\{\sigma\}} \bar{f} (Q,\{p_{j}\},\{m_{k}\};\{%
\sigma_{l}\})  \nonumber \\
&&= \int d^{n}Q {\bf L}_{\{\sigma\}} \left ({\partial}_{p_{j}} \right
)^{\omega}\bar{f} (Q,\{p_{j}\},\{m_{k}\};\{\sigma_{l}\}) = {\bf L}%
_{\{\sigma\}} \int d^{n}Q \left ({\partial}_{p_{j}} \right )^{\omega}\bar{f}
(Q,\{p_{j}\},\{m_{k}\};\{\sigma_{l}\})  \nonumber \\
&&={\bf L}_{\{\sigma\}} \left ({\partial}_{p_{j}} \right)^{\omega}
\overline{%
\Gamma} (\{p_{j}\},\{m_{k}\};\{\sigma_{l}\}) = \left ( {\partial}_{p_{j}}
\right )^{\omega} \Gamma^{0} (\{p_{j}\},\{m_{k}\}).
\end{eqnarray}
The second and the fifth steps follow from the commutativity of the two
operations $\left ({\partial}_{p_{j}} \right )^{\omega}$ and ${\bf L}%
_{\{\sigma\}}$ as they act on different arguments, the third step is due to
the existence of $\int d^{n}Q \left({\partial}_{p_{j}} \right )^{\omega}
f(Q,...)$ and the fourth is justified from the existence of $\int d^{n}Q
\bar{f} (Q,...;\{\sigma_{l}\}) ( = \overline {\Gamma} (...;\{\sigma_{l}\}))$
by postulate.

It is clear that here the differentiation with respect to the external
momenta and its 'inverse'--indefinite integration with respect to the same
momenta play the role of the certain operations stated above.

The right hand side of Eq.(8) can be found now as the left end exists as a
nonpolynomial (nonlocal) function of external momenta and masses, i.e.,
denoting it as $\Gamma _{(\omega )}^0$, %%%eq(10)
\FL
\begin{equation}
\left( {\partial }_{p_j}\right) ^\omega \Gamma ^0(\{p_j\},\{m_k\})=\Gamma
_{(\omega )}^0(\{p_j\},\{m_k\}).
\end{equation}
To find $\Gamma ^0(\{p_j\},\{m_k\})$, we integrate both sides of Eq.(10)
with respect to the external momenta ''$\omega $'' times indefinitely to
arrive at the following expressions %%%%%Eq(11)
\FL
\begin{eqnarray}
\left( \int_{{p}}\right) ^\omega \left[ ({\partial }_{{p}})^\omega \Gamma
^0(\{p_j\},\{m_k\})\right] =\Gamma ^0(\{p_j\},\{m_k\})+N^\omega
(\{p_j\},\{c_\omega \}) =\Gamma _{npl}(\{p_j\},\{m_k\})+N^\omega
(\{p_j\},\{C_\omega \})
\end{eqnarray}
with $\{c_\omega \}$ and $\{C_\omega \}$ being arbitrary constant
coefficients of an $\omega -1$ order polynomial in external momenta $%
N^\omega $ and $\Gamma _{npl}(\{p_j\},\{m_k\})$ being a definite
nonpolynomial function of momenta and masses. Evidently $\Gamma
^0(\{p_j\},\{m_k\})$ is not uniquely determined within conventional QFTs at
this stage. That the true expression %%%%%eq(12)
\FL
\begin{equation}
\Gamma ^0(\{p_j\},\{m_k\})=\Gamma _{npl}(\{p_j\},\{m_k\})+N^\omega
(\{p_j\},\{\bar{c}_\omega \}),\ \ \ \bar{c}_\omega =C_\omega -c_\omega
\end{equation}
contains a definite polynomial part (unknown yet) implies that it should
have come from the LE limit operation on $\overline{\Gamma }%
(\{p_j\},\{m_k\};\{\sigma _l\})$ (see Eq.(5)) as the usual convolution
integration can not yield a polynomial part--an indication of the
incompleteness of the formalism of the QFTs.

We can take the above procedures as efforts for rectifying the ill-defined
FAs and ''represent'' the FAs with the expressions like the right hand side
of Eq.(11), i.e., %%%eq(13)
\FL
\begin{equation}
\int d^nQf(Q,\{p_j\},\{m_k\})>=<\Gamma _{npl}(\{p_j\},\{m_k\})+N^\omega
(\{p_j\},\{C_\omega \})
\end{equation}
with ''$>=<$'' indicating that left hand side is rectified as the right hand
side. That the ambiguities reside only in the local part means that the QFTs
are quite effective in the LE limit.

To find the $\{\bar{c}_{\omega}\}$'s in Eq.(12) we need inputs from the
physical properties of the system (such as symmetries, invariances and
unitarity) and a complete set of data from experiments \cite{CK,LL} (if we
can derive them from the underlying theory all these requirements would be
automatically fulfilled). In other words, all the ambiguities should be
fixed according to this principle. Similar approach had been adopted by
Llewellyn Smith to fix ambiguities on Lagrangian level by imposing high
energy symmetry, etc. on relevant quantities \cite{LL}. It is also the
physical reasoning followed by the conventional renormalization programs.

As we have seen, the $\bar{c}_\omega $'s arise in fact from the low energy
limit operation on the objects calculated in the underlying theory, they
should be uniquely defined up to possible equivalence. Different or
inequivalent choices of these constants simply correspond to different LE
theories (amount to being defined by different underlying theories). Since
different regularizations and/or renormalization conditions might correspond
to inequivalent choices of the constants, (the problem might be more severe
in nonperturbative cases \cite{QMDR}), they might lead to different LE
theories that even could not faithfully describe relevant low energy
physics. Although the underlying structures do not appear explicitly in the
LE formulations, they are not totally decoupled from the effective ones and
they 'stipulate' the effective sectors indirectly through the constants $\{%
\bar{c}\}$. In other words, these constants missing from the effective
Hamiltonian models are just what we need to define the quantum theory
completely in addition to the canonical parameters (masses and couplings).
Their missing in the Hamiltonian or Lagrangian level does not mean they do
not exist. As we will see in section IV, the former studies on the
self-adjoint extension \cite{QM} of some quantum mechanical Hamiltonians
just support our point of view here.

\section{Multi-loop Case}

Since the UV divergence would appear if one first take the limit before
doing loop momenta integrations, our strategy here is just to move the limit
operator ${\bf L}_{\{\sigma \}}$ across the loop integration operations in
such a way that no potential divergence is left over.

For any multi-loop graph $\Gamma $ (we will use the same symbol to denote
the graph and the associated FA if it is not confusing), we should again
start with the amplitude derived from the underlying theory, i.e., $%
\overline{\Gamma }(\ldots ;\{\sigma \})$ with the same graph structure. All
the internal lines and vertices are again understood to be given by the
underlying theory, characterized by the presence of the parameters $\{\sigma
\}$. Then the LE limit of $\overline{\Gamma }(\ldots ;\{\sigma \})$ produces
the definite constants $\{c^0\}$ that are unknown to us yet %%eq(14)
\FL
\begin{equation}
\Gamma ^0(\ldots ;\{c^0\})={\bf L}_{\{\sigma \}}\overline{\Gamma }(\ldots
;\{\sigma \})={\bf L}_{\{\sigma \}}\int \prod_ld^nl{\bar{f}}_\Gamma
(\{l\},\ldots ;\{\sigma \})
\end{equation}
where $\bar{f}_\Gamma (\{l\},\ldots ;\{\sigma \})$ denotes the integrand
obtained from the underlying theory corresponding to the graph $\Gamma $ and
the dots refer to the LE parameters like external momenta, mass parameters
and coupling constants. Other symbols are self-evident. We will use in the
following $\omega _\gamma -1$ to denote the overall divergence index \cite
{Shirkov} for any graph $\gamma $ and $\{l\}$ to represent the internal
momenta and all the partial differentiation operators and their 'inverse'
($%
\partial _{\omega _\gamma }^{-1}$) act upon the momenta only 'external' to
the very internal integration of the graph under consideration. In certain
cases, masses are appropriate or efficient external parameters to work with.

If the graph is totally convergent, then $\Gamma ^0$ contains no UV
ambiguity and the limit operation can cross all the internal integrations to
act upon the integrand to yield the product of the propagators and vertices
given by the present QFTs. But once there is any potential UV
ill-definedness associated with any internal integration, one must proceed
in the following way (suppose that a graph $\Gamma $ contains at least an
overall divergence) %%eq(15)
\FL
\begin{eqnarray}
&&\Gamma ^0(\ldots ;\{c_i^0\})={\bf L}_{\{\sigma \}}\int \prod d^nl\bar{f}%
_\Gamma (\{l\},\ldots ;\{\sigma \})  \nonumber \\
&\Rightarrow &\partial _{\omega _\Gamma }^{-1}{\bf L}_{\{\sigma \}}\int
\prod d^nl\partial ^{\omega _\Gamma }\bar{f}_\Gamma (\{l\},\ldots ;\{\sigma
\})=\sum_{\{\gamma \}=\partial ^{\omega _\Gamma }\Gamma }\partial _{\omega
_\Gamma }^{-1}{\bf L}_{\{\sigma \}}\int \prod d^nl\bar{f}_\gamma
(\{l\},\ldots ;\{\sigma \}).
\end{eqnarray}
The differentiation with respect to the external parameters will give rise
to a sum of graphs $\{\gamma \}$ (without overall divergence) from the
original graph $\Gamma $. (Note that any overlapping divergence will be
killed by the differentiation operation, only non-overlapping divergences
remain, i.e., the overlapping divergences are disentangled \cite{CK}). If
there is no more ill-definedness (in any subgraph), one can move the limit
operator across all the internal integrations to act directly upon the
integrands $\bar{f}_\gamma (\{l\},\ldots ;\{\sigma \})$ just like in the
totally convergent graph case. Then one can carry out all the loop
integrations without any trouble for each graph $\gamma $ and then sum them
up and finally apply the 'inverse' operator with respect to the parameters
(usually momenta) external to the graph $\Gamma $ (and each $\gamma $).

Now suppose there are still some ill-defined subgraphs in each $\gamma $. In
this case, each graph in the set ${\partial }^{\omega _\Gamma }\Gamma $ can
be expressed as a 'product' of disconnected divergent (at least
superficially divergent) subgraphs (each subgraph itself may contain
overlapping divergences). The LE limit operator crossed all the other parts
and stopped before the divergent subgraphs. In formula, for each graph $%
\gamma $,it is %%%Eq(16)
\FL
\begin{eqnarray}
&&\partial _{\omega _\Gamma }^{-1}{\bf L}_{\{\sigma \}}\int \prod
d^nl\bar{f}%
_\gamma (\{l\},\ldots ;\{\sigma \})=\partial _{\omega _\Gamma }^{-1}\left\{
\int \prod d^n\overline{l^{\prime }}g_{\gamma /{[\gamma ^{\prime }]}}(\{%
\overline{l^{\prime }}\},\ldots ){\bf L}_{\{\sigma \}}\prod_{\gamma
_j^{\prime }}\int \prod_{i\epsilon \gamma
_j^{\prime }}d^nl_i^{\prime }{\bar{%
f}}_{\gamma :\gamma _j^{\prime }}(\{l^{\prime }\}_{\gamma _j^{\prime
}},\ldots ;\{\sigma \})\right\}   \nonumber \\
&=&\partial _{\omega _\Gamma }^{-1}\left\{ \int \prod d^n\overline{l^{\prime
}}g_{\gamma /{[\gamma ^{\prime }]}}(\{\overline{l^{\prime }}\},\ldots
)\prod_{\gamma _j^{\prime }}\left[ {\bf L}_{\{\sigma
\}}{\overline{\Gamma }}%
_{\gamma :\gamma _j^{\prime }}(\ldots ;\{\sigma \})\right] \right\} , \\
&&(\bigcup_{\gamma _j^{\prime }}\{l^{\prime }\}_{\gamma _j^{\prime
}})\bigcup \{\overline{l^{\prime }}\}=\{l\},\ \ \ \ \ [\gamma ^{\prime
}]\bigcup \gamma /{[\gamma ^{\prime }]}=\gamma ,\ \ \ [\gamma ^{\prime
}]=\prod_j\gamma _j^{\prime },\gamma _j^{\prime }\bigcap \gamma _k^{\prime
}=0,\ \ j\neq k,
\end{eqnarray}
where all the dots in the expressions refer to the parameters 'external' to
the loop integrations for the subgraphs (i.e., to $\gamma _j^{\prime }$%
)--they contain the external parameters for the original graph $\Gamma $
(also for all the graphs in $\{\gamma \}$) {\it and} the internal momenta in
the set $\{\overline{l^{\prime }}\}$. $\overline{\Gamma }_{\gamma :\gamma
_j^{\prime }}$ refers to amplitude derived from the underlying theory that
corresponds to each subgraph $\gamma _j^{\prime }$ contained in $\gamma $.
{\it Since some loop momenta are 'external' to certain subgraphs, one can
not first carry out these loop integrations before the ill-defined subgraphs
are treated. } This is in sheer contrast to the totally convergent graphs
where the loop integration order does not matter.

As the ill-defined subgraphs in $[\gamma ^{\prime }]$ are disconnected with
each other, we now treat each of them separately as a new 'total' graph just
like what we have done with the total graph $\Gamma $ starting from Eq.(14).
Then we go through the procedures from Eq.(14) to Eq.(17) till we encounter
new disconnected and ill-defined subgraphs that are in turn to be treated as
before. Finally, we will go to the smallest subgraphs that are completely
convergent. Now we can finally remove the LE limit operator to get the
integrands totally expressed with propagators and vertices given by the
effective theories and we can begin to perform all the loop integrations
with insertions of various pairs of $\partial _{\omega _\gamma }^{-1}$ and
$%
\partial ^{\omega _\gamma }$ and there will appear a natural order of
integration due to our procedures above: the smallest subgaphs are
integrated first, then the ''larger'' subgraphs with their subgraphs done
already, and then the still ''larger'' subgraphs and so on, till all
integrations are done. [It is worthwhile to note that at each level of the
subgraphs, the loop integrations are guaranteed to be convergent due to
Weinberg's theorem \cite{weinth}].

The resulting expression will be a definite nonlocal functions plus nonlocal
ambiguities (due to subgraph ill-definedness) and local ambiguities if $%
\Gamma$ is suffering from overall divergence, %Eq(18)
\FL
\begin{eqnarray}
&&\Gamma^0(\ldots; \{c^0\}) \Rightarrow \Gamma(\ldots;\{C\})
=\Gamma^{npl}_{0}(\ldots)+ \Gamma^{npl}_{1}(\ldots;\{C^{\prime}\})+
N^{\omega_{\Gamma}}(\ldots;\{\bar{C^{\prime}}\}), \\
&&\{C^{\prime}\} \bigcup \{\bar{C^{\prime}}\}=\{C\}.
\end{eqnarray}
Here again we used $N^{\omega_{\Gamma}}$ to denote the polynomial containing
the ambiguities ($\{\bar{C^{\prime}}\}$) appearing due to the overall
divergence. Others are nonlocal functions. Different from the single loop
case, there are nonlocal ambiguities in this multiloop graph suffering from
subgraph divergences (as evident from our treatment) in addition to the
nonlocal definite part and the local ambiguous part. The result we obtained
(%
$\Gamma (\ldots;\{C\})$) is not what we are really after ($%
\Gamma^0(\ldots;\{c^0\})$), but that is the best we can do with the present
QFT.

Now some remarks are in order:

{\bf A}. It is evident that overlapping divergences are just automatically
resolved in our proposal, because the differentiation operators just remove
the overlapping ill-definedness by 'inserting' internal lines and vertices
to reduce the overall divergence. Thus one need not worry about them any
more. This is the utility derived from the differentiation with respect to
external parameters (momenta, masses or other massive parameters that might
appear in the LE propagators) \cite{CK}.

{\bf B}. The amplitudes given by the underlying theory should be invariant
under any linear transformations of the internal integration variables. In
our treatment of the ill-defined graphs, since every loop integration
actually performed is convergent, these transformations do not alter the
results of the loop integrations. Due to the 'inverse' operations, these
linear transformations would at most change the polynomial part. But that
does not matter at all. This observation implies that one should not worry
about the variable shifting and routing of the external momenta that belong
to the transformations just described if he has noted the ambiguous part.

{\bf B1}. An immediate corollary to this observation is that, the chiral
anomaly, which is conventionally interpreted as due to the variable shifting
in relevant linearly divergent amplitude, must have been due to other
definite properties. Otherwise, if it were totally due to the local
ambiguities, one can well remove them away by choosing appropriate
definitions of the constants (or appropriate renormalization conditions).
Our direct calculation shows that \cite{JF1}, one kind of definite rational
terms (independent of masses) originated the chiral anomaly. Since they are
nonlocal and unambiguous, one can not attribute them simply as UV effects.
The trace anomaly is also shown to be originated by such kind of rational
terms \cite{JF1}. To our best knowledge, this nontrivial structure
(independent of the UV ambiguities) has never been noted before in the old
renormalization framework.

{\bf B2}. Another utility derived from the observation is that, one can
choose the routes of flows of the external momenta to be as simple as
possible to make the treatments of an ill-defined multi-loop amplitude as
easy as possible. For the single loop cases, sometimes one may only focus on
the parts of the amplitude that are really divergent. This may yield fewer
ambiguities.

{\bf C}. With the above deduction, one can easily see that the results of
any regularization and/or renormalization scheme can be readily reproduced
by corresponding choices of the constants. That is, our proposal can lead to
a universal formulation for all the regularization and renormalization
schemes at least in the perturbative framework. If one wish to pursue
calculation efficiency in the first place, one can choose a regularization
scheme that saves labor and then replace all the divergent expressions with
an ambiguous polynomial in external parameters to arrive at the same
expressions that can be obtained in the procedures described above.
Conventionally one has to check whether the symmetry properties of a scheme
would affect the main body of a theory and/or works very hard in order to
setup a scheme as consistently as possible. Our proposal spares the labor of
checking every corner of a scheme and just make use of its efficiency.

For the ambiguities, as we have discussed in the section II, we may first
impose some novel symmetries and invariances on the amplitudes to reduce the
ambiguities to certain degree, then one has to resort to the experimental
physics data, a strategy effectively employed in the conventional
renormalization approaches. The W-T identities have served as constraints in
the conventional renormalization schemes and saved a lot of labor of
calculation. They are just certain kinds of graphical relations among the
Feynman graphs. So, in the Feynman graph language, we can derive constraints
upon the ambiguities from the graph structures \cite{YYY}.

Moreover, it is worthwhile to note that our strategy works in any kind of
models as long as they are physically effective and consistent. That is, it
applies efficiently to the unrenormalizable theories as well as the
renormalizable ones in the conventional terminology. It would be interesting
to integrate our approach with the BV anti-field formalism of QFT \cite{BV}
that has been used to deal with the unrenormalizable theories quite recently
\cite{gom}. In contrast, we do not need any counterterm in our strategy for
whatever kind of unrenormalizable models simply because we do not encounter
any divergence, the only trouble is to fix the ambiguities..

As we have pointed out, the Feynman Amplitudes or the 1PI functions are
generally parametrized by more than one constants (I will refer to them as
'agent constants') in addition to the phenomenological ones. If the changes
in the radiative constants could be completely compensated by that in the
phenomenological ones (which is only possible for rather special kind of
models), then we might implement a redefinition invariance of the
phenomenological constants or parameters for the FAs like in the RG case.
Recently we have rederived the Callan-Symanzik like equations governing the
scale transformation behaviors of QFTs without divergence and bare
parameters. The influences of the underlying structures effected through the
agent constants are clearly shown and these agent constants complete the
harmony of the scale transformation of the effective QFTs in the sense
discussed in Ref. \cite{Scale}. Moreover, the new equations improved the
conventional ones in quite some aspects.

\section{Nonperturbative Examples}

Iit is obvious that our proposal works in principle for any model, whether
it is a QFT or not. The key observation that the UV ill-definedness is
caused by illegitimate order of 'operations' is valid in both perturbative
and the nonperturbative contexts. That is to say, our simple and natural
strategy should also be applicable to nonperturbative case.

Recently, the cutoff and Dimensional regularizations are compared in
nonperturbative context \cite{QMDR} in the hot topic of applying the idea of
effective field theory method (EFT \cite{EFFT}) to LE nuclear physics
following Weinberg's suggestion \cite{WeinEFT}. The framework is a
non-relativistic quantum mechanics with Delta-potentials, which is ill
defined in the short-distance. According to our discussions above, such
ill-definedness means that the effective LE models must have failed in the
higher energy end. Then it is illegitimate to simply work with the
propagators and vertices (or Green functions and potentials) given by such
models. Great care must also be taken with respect to the regularization
effects. Thus the inequivalence between the cutoff scheme and Dimensional
regularization exhibited in Ref. \cite{QMDR} well evidenced the correctness
of our arguments given above.

Now let us try to treat the problem within our proposal. Generally, the
Lippmann-Schwinger equation for $T$-Matrix in the simple two-body problems
reads (we follow the notation conventions of Ref. \cite{QMDR}) %%%eq(20)
\FL
\begin{equation}
T(p^{\prime},p; E)=V(p^{\prime},p)+ \int \displaystyle\frac{d^d k}{(2\pi)^d}
V(p^{\prime},k ) \displaystyle\frac{1}{E^{+}-k^2/(2\mu)} T(k,p; E),
\end{equation}
where $E^{+}$ is $E + i \epsilon$, with $E$ non-negative, and $\mu$ denotes
the reduced mass in the two-body problem. In our point of view, this
equation is not well-defined and should be written as the LE limit of that
derived from the well-defined fundamental underlying theory which is
unavailable to us by now. [The underlying parameters will be always denoted
as $\{\sigma\}$. For different problems or different LE ranges, the contents
may differ.] So in our language, Eq.(20) should be rewritten as %%eq(21)
\FL
\begin{eqnarray}
&&T(p^{\prime},p; E ; \{\sigma\}) = V(p^{\prime},p ; \{\sigma\}) + \int %
\displaystyle\frac{d^d k}{(2\pi)^2} V(p^{\prime},k ; \{\sigma\})
G(E^{+}-k^2/(2\mu); \{\sigma\}) T(k,p;E; \{\sigma\}), \\
&&V(p^{\prime},p) \equiv {\bf L}_{\{\sigma\}} V(p^{\prime},p ; \{\sigma\}),
\displaystyle\frac{1}{E^{+}-k^2/(2\mu)}\equiv {\bf L}_{\{\sigma\}}
G(E^{+}-k^2/(2\mu); \{\sigma\}).
\end{eqnarray}
Eq.(21) is now well-defined in the underlying theory. Thus Eq.(20) is
correct only when there is no UV infinities (there is no IR problem in the
following discussions for the Delta-potential problem) so that the LE limit
operator can cross the internal momentum integration (summation over
intermediate states). Otherwise we have to find a legitimate way to let the
LE limit operator cross the internal momenta integration.

In the case of Delta-potential, $V(p^{\prime },p)=C$, but the $V(\ldots
;\{\sigma \})$ is generally a nonlocal potential before the LE limit is
taken. To be rigorous, we write formally %%eq(23)
\FL
\begin{eqnarray}
T(p^{\prime },p;E;\{c^0\}) &=&C+{\bf L}_{\{\sigma \}}\left\{ \int %
\displaystyle\frac{d^dk}{(2\pi )^d}V(p^{\prime },k;\{\sigma \})
G(E^{+}-k^2/(2\mu );\{\sigma \})T(p^{\prime },p;E;\{\sigma \})\right\} ,
\end{eqnarray}
here it is not generally legitimate to move the $V(\ldots ;\{\sigma \})$ out
of the integration to be directly subject to the LE limit operator--which is
exactly what was done in the conventional calculations (with only the
propagator regularized)--and it is definitely illegitimate to apply the LE
limit operator to all the other objects before the integration is done.
Thus, in principle, even when the LE potential is local (of course $V(\ldots
;\{\sigma \})$ is nonlocal), it might be dangerous to simply reduce Eq.(21)
to an algebraic one. Only when the ill-definedness is mainly caused by $%
1/(E^{+}-k^2/(2\mu ))$ (i.e., it differs greatly from $G(\ldots ;\{\sigma
\}) $ in the UV region where $V(\ldots )$ differs less from $V(\ldots
;\{\sigma \})$), could we pull out the true potential to subject it directly
to the action of the LE limit operator. In other words, to put Eq.(21) (a
correct formulation for Eq.(20)) or Eq. (23) into an algebraic one requires
quite nontrivial properties of the potential and the propagator. To focus on
the main point, we temporarily assume this condition is satisfied, then we
have the well-defined form of the algebraic equation for the $T-$matrix
(which is now parametrized by the new constants $\{c^0\}$ from the LE limit
in addition to $E$), %eq(24)
\FL
\begin{equation}
\displaystyle\frac 1{T^{on}(E;\{c^0\})}=\displaystyle\frac 1C-I(E;\{c^0\}),
\end{equation}
with %%eq(25)
\FL
\begin{equation}
I(E;\{c^0\})={\bf L}{\{\sigma \}}\int \displaystyle\frac{d^dk}{(2\pi )^d}%
G(E^{+}-k^2/(2\mu );\{\sigma \}).
\end{equation}
Now we can employ the technique described in sections II and III to
calculate the integrals, i.e., first differentiate $G(E^{+}-\ldots
;\ldots )$
with respect to $E^{+}$ (which is the 'external' parameter in the integral)
for appropriate times, secondly perform the LE limit legitimately and carry
out the integral thus obtained, finally do the 'inverse' operation with
respect to $E$ and we find the followings (note that here one
differentiation with respect to $E$ reduces the divergence degree by two)
%%%eq(26,27)
\FL
\begin{eqnarray}
&&I_{d;odd}(E;\{c^0\})\Rightarrow I_{d;o}(E;\{c^{\prime }\})
=-i\displaystyle%
\frac{2\mu {\pi }^{d/2+1}}{\Gamma (d/2)(2\pi )^d}(2\mu
E)^{d/2-1}+N^{[(d-1)/2]}(2\mu E;\{c^{\prime }\}); \\
&&I_{d;even}(E;\{c^0\})\Rightarrow I_{d;e}(E;\{c^{\prime }\})
=\displaystyle%
\frac{2\mu \pi ^{d/2}}{\Gamma (d/2)(2\pi )^d}(2\mu E)^{d/2-1}\ln (2\mu
E/c_0^{\prime })+N^{[d/2]}(2\mu E;\{c^{\prime }\})
\end{eqnarray}
with $\{c^{\prime }\}$ being arbitrary constants--the ambiguities. These
expressions can again be viewed as universal parametrization and compared
with that given in cutoff regularization and dimensional regularization
schemes (C.f. Ref \cite{QMDR}) with the latter ones as special cases.

In terms of the ambiguous (but {\it finite }) integrals given by Eq.(26,27),
the $T-$matrix is now parametrized by $\{c^{\prime}\}$ in addition to $E$
like %%eq(28)
\FL
\begin{equation}
\displaystyle\frac{1}{T^{on}(E;\{c^{\prime}\})} = \displaystyle\frac{1}{C} -
I_{d;\ldots}(E;\{c^{\prime}\}).
\end{equation}
Again we need to fix the constants $\{c^{\prime}\}$ rather than to
renormalize the interaction constant $C$.

It is easy to see that following the normalization condition of Ref. \cite
{QMDR}, we can reproduce the result derived by Weinberg \cite{WeinEFT} in
two or three dimensional space-time. However, there seems to be no necessary
constraints on the phenomenological constant $C$ as it is physical rather
than ' bare' in our proposal. Thus, this LE framework \'{a} la Weinberg
works equally well for both the attractive interactions and the repulsive
ones in our approach , contrary to the conclusions that EFT framework failed
in the repulsive cases where the LE models are believed to be trivial \cite
{Beg}. The nontriviality of the Delta-potential dynamics has also been
investigated by Jackiw \cite{Jackiw}.

This problem can also be examined from another angle. Jackiw had already
pointed out \cite{Jackiw} that the Hamiltonians for such models are not
automatically Hermitian and need self-adjoint extension. This has already
been dealt with by mathematicians in the operator theory and has also been
extensively discussed in a number of approaches (please refer to \cite
{Albeverio} for a comprehensive list of the references). The key point is,
in such cases, the self-adjointness of the Hamiltonian is never an automatic
property. That is, in contrast to the normal case, the contact potential
problem \'{a} la Schr\"{o}dinger equation is UV ill-defined. The resolution
of the problem gives rise to a family of self-adjoint extensions of the
original Hamiltonian operator parametrized by an additional constant, which
upon different choices leads to different or inequivalent (LE) physics \cite
{QM}. {\it This additional 'family' parameter is just the constant that will
surely be predicted from the LE limit operation in our proposal}, as has
been emphasized in section II and III. As a matter of fact, there is an
approach that is quite the same as ours in spirit, the one based on
resolvent formalism \cite{Grossmann} where an important object is defined
through an equation in which it appeared in a form differentiated with
respect to the 'external' parameter--the resolvent variable (energy). Thus
this important object is only defined up to an additional parameter (the
family parameter in operator theory approach) which is to be determined by
other input, just like in our proposal.

I would like to mention a recent calculation \cite{GEP} of Higgs masses in
nonperturbative context employing our proposal. The results thus obtained
are neat and clear, in contrast to that performed within the old
renormalization framework (see the references in \cite{GEP}). Especially,
the physical pictures are different from that using the old renormalization,
which is easy to see from the discussions above.

One could expect that great ease can be found in employing our proposal or
its equivalents (in any form known or unknown) in his/her studies in the
nonperturbative contexts and the outcome would be quite different and
significant. Moreover, within our approach, those phenomenologically
oriented models which are unrenormalizable in the usual renormalization
schemes could become quite tamed. One can test it with the NJL model and
chiral perturbation theory \cite{CPT} and with gravity \cite{Dono}. We will
further apply our strategy to more concrete calculations and develop more
rigorous framework of this strategy in the future.

\section{Discussions and summary}

For a complete representation of the world, we should expect that the
underlying theory is also well defined in the IR sector. (Our discussions in
the introduction about the spectra are partial as we deliberately omitted
the IR issues to focus on the UV structures.) It is conceivable that the
phenomenological models give wrong information of the IR end spectra
signaled by the unphysical IR infinities.

The underlying theory, 'postulated' here, if exists, should contain all the
nontrivial UV and IR structural information lost in the effective theories.
Then an interesting scenario dawns upon us: for each effective model
dominating certain energy range (say, theory $T_{mid}$), there should exist
two other effective models (or sectors) that are most adjacent to this model
from the IR end and UV end respectively (say, $T_{IR}$ and $T_{UV}$). Then
it is imaginable that the phenomenological parameters in $T_{IR}$ and/or $%
T_{UV}$ would at least quite nontrivially improve the IR and/or UV behaviors
of the theory $T_{mid}$. While on the other hand, the $T_{mid}$ contains
what $T_{IR}$ (or $T_{UV}$) needs to improve its UV (or IR) behaviors. Put
it another way, the active and 'elementary' modes or fields in $T_{IR}$ will
break up in $T_{mid}$ and give way to the new 'elementary' modes active in
$%
T_{mid}$. Similarly, the 'elementary' modes in $T_{mid}$ will go
'hibernating' as the energy goes down while 'new' elementary modes 'emerge'
to dominate spectra in $T_{IR}$. The relation between the elementary modes
in $T_{mid}$ and $T_{UV}$ is in principle just like that between those in $%
T_{IR}$ and $T_{mid}$. Of course, there may be modes active in several
successive effective models, some may even be active and stable through all
energy levels---the 'fossil' modes or fields we mentioned in the
introduction. Evidently, the information about those 'elementary' modes in
$%
T_{IR}$ and $T_{UV}$ missing from $T_{mid}$ (i.e., missing from the
effective spectrum given by $T_{mid}$) can contribute to improve the IR and
UV behavior of the latter.

Of course the author does not have a solid idea for the answer right now.
Thus we only indicate some interesting but plausible picture about the
resolution of the unphysical IR problem or the reformulation of the IR end
of a theory suffered from IR problems like QCD.

From our proposal, the 'elementary'commutator for a field and its conjugate,
if calculated (or formulated) from the underlying theory, must have been at
least a nonlocal function(al) parametrized by the underlying parameters of
the underlying theory and must have been closely related with the
gravitational interaction and perhaps new fundamental ones, rather than a
highly abstract Dirac delta function containing least information. In a
sense, the incompleteness of the present QFTs or their ill-definedness is
inherent in the present quantization procedure whose most elementary
technical building block is the Dirac delta function (called as distribution
by mathematicians) that is {\it extremely singular and oversimplified in the
UV ends.} That the distribution theory works necessarily with test function
space or appropriate measure, if viewed from physical angle, is equivalent
to that we need more 'fundamental structures' in order for some singular
functions to make sense, i.e., a necessity of introducing underlying theory
or its artificial substitute--regularization.

In a short summary, we discussed the strategy recently proposed by the
author and the important consequences following from it. The proposal could
overcome many typical difficulties and shortcomings associated with old
regularization and renormalization framework and can be applied in principle
to any UV ill-defined quantum theories.

\section*{Acknowledgement}

The author is very grateful to Professor F. V. Tkachov for his encouragement
and enlightening remarks.

\end{document}